\documentstyle[12pt]{article}
\def\heet#1{#1\kern -1ex /}

\begin{document}

\begin{center}

{\Large\bf
Free totally (anti)symmetric massless fermionic

\medskip
fields in d-dimensional anti-de Sitter space}

\vspace{2cm}
R.~R.~Metsaev

\vspace{1cm}
Department of Theoretical Physics, P.~N.~Lebedev Physical
Institute, Leninsky prospect 53, 117924, Moscow, Russia

\vspace{3cm}

{\bf Abstract}
\end{center}

\noindent Free massless fermionic fields of arbitrary spins $s>0$
corresponding to totally (anti)sym\-metric
tensor-spinor representations of the $SO(d-1)$ compact subgroup
and in $d$-dimensional anti-de Sitter space are
investigated.  We propose the free equations of motion,
subsidiary conditions and corresponding gauge transformations
for such fields.  The equations obtained are used to derive the
lowest energy values for the above-mentioned representations. A
new representation for equations of motion and gauge
transformations in terms of generators of anti-de Sitter group
$SO(d-1,2)$ is found.
It is demonstrated that in contrast to the symmetric case
the gauge parameter of the antisymmetric massless field is
also a massless field.

\newpage

The long term motivation of our investigation of higher-spin
massless field theory in the anti-de Sitter space of higher
dimensions is to study the equations of interacting gauge fields
of all spins suggested in \cite{VAS2}. Because these equations
are formulated in terms of wavefunctions which depend on usual
spacetime coordinates and certain twistor variables it is not
clear immediately what kind of fields they describe.
At present because of \cite{VAS2}, it is known that in four
dimensions ($d=4$) they describe unitary dynamics of massless
fields of all spins in anti-de Sitter space.  It is
expected that for the case of higher dimensions $d>4$ they also
describe unitary dynamics. The case of half-integral  spins,
presented here is a necessary step in our study of massless
higher spins of all spins for arbitrary $d$.
For $d=4$ the integral spins have been studied in \cite{F1}
and for arbitrary $d$ in \cite{MET1}.
We also hope that our results may be of wider interest,
especially to anti-de Sitter supergravity theories.

First let us  formulate the main problem that we will solve in
this letter. The positive-energy lowest weight
irreducible representation of $so(d-1,2)$ algebra denoted
as $D(E_0, {\bf m})$, is defined by  $E_0$, the lowest eigenvalue
of the energy operator, and by ${\bf m}=(m_1,\ldots m_\nu)$,
$\nu=[\frac{d-1}{2}]$, which is the weight of the $so(d-1)$
representation.  For the case of the four dimensional
anti-de Sitter space ($d=4$) it has been discovered
\cite{FH} that for fermionic massless fields $E_0=s+3/2$,
${\bf m}=(m_1=s+1/2)$ (see also \cite{FF},\cite{BRFR}).
In the present work we generalize this result  for the case of
arbitrary $d$ and for the fermionic fields corresponding to the
following two representations of the $so(d-1)$ algebra.  The
first representation has ${\bf m}_{sym}=(s+1/2,1/2,\ldots,1/2)$,
which is a totally symmetric representation and the second has
${\bf m}_{as}=(3/2,\ldots,3/2,1/2,\ldots,1/2)$ (where the 3/2
occurs $s$ times in this sequence) which is a totally
antisymmetric one.   For gauge fields labeled by ${\bf m}_{sym}$
and ${\bf m}_{as}$  the integer $s$ satisfies $s>0$ and
$1<s\leq \nu$, respectively.  In this paper we restrict
ourselves to the case of even $d$.

Now let us describe our conventions and notation. We
describe the anti-de Sitter space as a hyperboloid
$\eta_{AB}^{\vphantom{5pt}} y^A y^B=1$ in a $d+1$- dimensional
pseudo-Euclidean space with metric tensor
$\eta^{\vphantom{5pt}}_{AB}
=(+,-,\ldots,-,+)$, $A,B=0,1,\ldots,d-1,d+1$.
To simplify our expressions we will drop the metric tensor
$\eta_{AB}^{\vphantom{5pt}}$ in scalar products.

The generators of the $SO(d-1,2)$ group $J^{AB}$
satisfy the commutation relations

$$
[J^{AB},J^{CD}]={\rm i}\eta^{BC}J^{AD}
+\hbox{three terms}\,.
$$
We split $J^{AB}$ into an orbital part $L^{AB}$ and a spin part
$M^{AB}$: $J^{AB}=L^{AB}+M^{AB}$. The realization of
$L^{AB}$ in terms of a differential operator defined on the
hyperboloid $y^Ay^A=1$ is

$$
L^{AB}=y^A p^B-y^B p^A\,,
\qquad
p^A\equiv {\rm i}\theta^{AB}\frac{\partial}{\partial y^B}\,,
\qquad
\theta^{AB}\equiv \eta^{AB}-y^Ay^B\,,
$$
where $p^A$ is a momentum operator.
The $p^A$ has the properties $y^A p^A=0$,
$p^A y^A ={\rm i}d$ and satisfies the commutation relations
$[p^A,y^B]={\rm i}\theta^{AB}$ and $[p^A,p^B]={\rm i} L^{AB}$.
A form for $M^{AB}$ depends on the
realization of the representations. We will use the tensor
realization of representations. As the carriers for
$D(E_0,{\bf m})$ we use totally (anti)symmetric
tensor-spinor fields $\Psi_{as(sym)}^{A_1\ldots A_s}$
over the hyperboloid $y^Ay^A=1$. Note we do not show explicitly
spinor index of $\Psi$. Sometimes it
will be convenient to use the generating function

\begin{equation}\label{genfun}
|\Psi_{as(sym)}\rangle\equiv
(s!)^{-1}a_{as(sym)}^{A_1}\ldots a_{as(sym)}^{A_s}
\Psi_{as(sym)}^{A_1\ldots A_s}|0\rangle\,,
\end{equation}
where $|0\rangle$ is the Fock vacuum (i.e.
$\bar{a}^A|0\rangle=0$) and $a^A$ and $\bar{a}^A$ satisfy the
(anti)com\-mu\-ta\-tion relations

$$
\{\bar{a}^A_{as},\, a^B_{as}\}
=-\eta^{AB}\,,
\qquad
[\bar{a}^A_{sym},\, a^B_{sym}]=-\eta^{AB}\,,
$$
and $\{a_{as}^A,a_{as}^B\}=0$, $[a_{sym}^A,a_{sym}^B]=0$.
The subscripts $as$ and $sym$ will often be dropped in the
following when there is no ambiguity.
For realizations given by (\ref{genfun}), $M^{AB}$ has the
form

$$
M^{AB}=M_b^{AB}+\sigma^{AB}\,,
$$
where

$$
M_b^{AB}=-{\rm i}(a^A\bar{a}^B-a^B\bar{a}^A)\,,
\qquad
\sigma^{AB}=\frac{{\rm i}}{4}(\gamma^A\gamma^B
-\gamma^B\gamma^A)
$$
and $\gamma^A$ are gamma matrices:
${}\{\gamma^A,\gamma^B\}=2\eta^{AB}$.

Being the carriers for $D(E_0,{\bf m})$, the fields $\Psi$
satisfy the equations

\begin{equation}\label{eqmot1}
(Q-\langle Q\rangle)|\Psi\rangle=0\,,
\end{equation}
and allow the following subsidiary covariant constraints:

\begin{eqnarray}
\label{trans}
&&y^{A_1}\Psi^{A_1\ldots A_s} =0\qquad (\hbox{transversality})\,,
\\
\label{gtrans}
&&\gamma^{A_1}\Psi^{A_1\ldots A_s}=0 \qquad
\hbox{}
\\
\label{div}
&&p^{A_1}\Psi^{A_1\ldots A_s} =0\qquad
(\hbox{divergencelessness})\,,
\\
\label{trac}
&&\Psi_{sym}^{AAA_3\ldots A_s}=0\qquad\,\,
(\hbox{tracelessness})\,.
\end{eqnarray}
In equation (\ref{eqmot1}) $Q$ is the second-order Casimir operator of
the $so(d-1,2)$ algebra

$$
Q=\frac{1}{2}J^{AB}J^{AB}\,,
$$
while $\langle Q\rangle$ are its eigenvalues for
totally (anti)symmetric representations

\begin{eqnarray*}
&&\langle Q_{as}\rangle
=E_0(E_0+1-d)+s(d-s)+\frac{1}{8}(d-1)(d-2)\,,
\\
&&\langle Q_{sym}\rangle
=E_0(E_0+1-d)+s(s+d-2)+\frac{1}{8}(d-1)(d-2)\,.
\end{eqnarray*}
The $\langle Q\rangle$ can be calculated according to
the well known procedure \cite{PERPOP}. $E_0$ is the lowest eigenvalue
of $J^{0d+1}$.
The expression for $Q$ can be rewritten as

$$
Q=p^2+L^{AB}M^{AB}+
\frac{1}{2}M^{AB}M^{AB}\,,
\qquad
p^2\equiv p^Ap^A\,.
$$
Now taking into account the easily derived equalities
$\sigma^{AB}\sigma^{AB}=d(d+1)/4$ and

$$
L^{AB}M_b^{AB}|\Psi\rangle=-2s|\Psi\rangle\,,
\qquad
L^{AB}\sigma^{AB}|\Psi\rangle
={\rm i}\heet{y}\heet{p}|\Psi\rangle\,,
\qquad
\sigma^{AB}M_b^{AB}|\Psi\rangle=s|\Psi\rangle\,,
$$
$$
M_b^{AB}M_b^{AB}|\Psi_{(as)sym}\rangle
=\left\{
\begin{array}{l}
2s(d+1-s)|\Psi_{as}\rangle \,,
\\ [7pt]
2s(s+d-1)|\Psi_{sym}\rangle\,,
\end{array}
\right.
$$
$\heet{p}^2=p^2+{\rm i}\heet{y}\heet{p}$,
where $\heet{p}\equiv \gamma^Ap^A$ and
$\heet{y}\equiv \gamma^A y^A$, $\{\heet{p},\heet{y}\}={\rm i}d$,
one can rewrite equation (\ref{eqmot1}) as

\begin{equation}\label{eqmot2}
(\heet{p}^2-m^2)|\Psi\rangle=0\,,
\end{equation}
where we introduce formally the ``mass'' term

\begin{equation}\label{masener}
m^2\equiv  (E_0+\frac{1}{2})(E_0+\frac{1}{2}-d)\,.
\end{equation}

To define $E_0$ corresponding to massless fields we should respect
gauge-invariant equations of motion for $\Psi$. For the
case of totally symmetric representations we could use the
interesting results of \cite{VAS} (see also
\cite{LOPVAS}), where the gauge-invariant equations were
formulated in terms of linear curvatures. However, we prefer
to use a simpler and as it seems to us more adequate procedure
to derive $E_0$.  We proceed in the following way.

First we write the most general gauge transformations for
$\Psi^{A_1\ldots A_s}$

\begin{equation}\label{gaugetr1}
\delta \Psi^{A_1\ldots A_s}
=\sigma(A)\left(
p^{A_1}\Lambda^{A_2\ldots A_s}
+y^{A_1}R^{A_2\ldots A_s}
+\gamma^{A_1}S^{A_2\ldots A_s}\right)
+\mbox{cyclic permutations}
\end{equation}
where $\Lambda^{A_2\ldots A_s}$, $R^{A_2\ldots A_s}$, and
$S^{A_2\ldots A_s}$ are the totally (anti)symmetric
tensor-spinor fields.
In equation (\ref{gaugetr1}) and below  `cyclic permutations'
indicates the terms which are obtainable by making cyclic
permutations of indices $A_1,\ldots, A_s$.
For totally symmetric representations
$\sigma^{\vphantom{5pt}}_{sym}(A)=1$ while for the
antisymmetric one we have
$\sigma{\vphantom{5pt}}_{as}(A)=1$
for even cyclic permutations of $(A_1\ldots A_s)$ and
$\sigma{\vphantom{5pt}}_{as}(A)=(-1)^{s-1}$ for odd
cyclic permutations of $(A_1\ldots A_s)$.
On the gauge parameter fields
$\Lambda^{\ldots}$, $R^{\ldots}$ and $S^{\ldots}$ we impose the
algebraic constraints like (\ref{trans})--(\ref{trac}):

\begin{eqnarray}
&&
\label{gptrans}
y^{A_2}\Lambda^{A_2\ldots A_s}
=y^{A_2}R^{A_2\ldots A_s}=y^{A_2}S^{A_2\ldots A_s}=0\,,
\\
&&
\gamma^{A_2}\Lambda^{A_2\ldots A_s}
=\gamma^{A_2}R^{A_2\ldots A_s}=\gamma^{A_2}S^{A_2\ldots A_s}
=0\,,
\\
&&
p^{A_2}\Lambda^{A_2\ldots A_s}
=p^{A_2}R^{A_2\ldots A_s}=p^{A_2}S^{A_2\ldots A_s}=0\,,
\\
\label{gptrac}
&&\Lambda_{sym}^{AAA_4\ldots A_s}
=R_{sym}^{AAA_4\ldots A_s}=S_{sym}^{AAA_4\ldots A_s}=0\,.
\end{eqnarray}

Secondly, we require that subsidiary conditions
(\ref{trans})--(\ref{trac}) and (\ref{eqmot2})  be invariant
with respect to the gauge transformation (\ref{gaugetr1}).
Note that due to (\ref{gptrans})--(\ref{gptrac})
the invariance requirement of (\ref{trac}) is already
satisfied.
From the invariance requirement
of (\ref{trans}) (i.e.  $y^{A_1} \delta\Psi^{A_1\ldots}=0$)
it follows

\begin{equation}\label{rsol}
R^{A_2\ldots A_s}
=\mp {\rm i}(s-1)\Lambda^{A_2\ldots A_s}
-\heet{y} S^{A_2\ldots A_s}\,.
\end{equation}
In equation (\ref{rsol}) and in the following upper and
lower signs correspond to antisymmetric and symmetric cases,
respectively.
Thus at this stage we have the gauge transformations

\begin{equation}\label{gaugetr2}
\delta \Psi^{A_1\ldots A_s}
=\sigma(A)\left(
(p^{A_1}\mp {\rm i}(s-1)y^{A_1})\Lambda^{A_2\ldots A_s}
+\theta^{A_1B}\gamma^B S^{A_2\ldots A_s}\right)
+\mbox{cyclic permutations}
\end{equation}
From the invariance requirement of
(\ref{gtrans}) with respect to (\ref{gaugetr2})
(i.e. $\gamma^{A_1}\delta\Psi^{A_1\ldots}=0$) we get

\begin{equation}\label{gammainv1}
(\heet{p}\mp {\rm i}(s-1)\heet{y})\Lambda^{A_2\ldots A_s}
+(d\mp (2s-2))S^{A_2\ldots A_s}=0\,.
\end{equation}
From the invariance requirement of (\ref{div})
(i.e. $p^{A_1}\delta\Psi^{A_1\ldots}=0$) we get

\begin{equation}\label{mominv}
\left(p^2-(s-1)(s-1\mp(d-1))\right)\Lambda^{A_2\ldots A_s}
+\left(\heet{p}\pm {\rm i}(s-1\mp d)\heet{y}\right)
S^{A_2\ldots A_s}=0\,.
\end{equation}
From the invariance requirement of (\ref{eqmot2})
(i.e. $(\heet{p}^2-m^2)\delta\Psi^{\ldots}=0$) we get

\begin{equation}\label{masval}
m_{as}^2=s(s-d)\,,
\qquad
m_{sym}^2=(s-2)(s+d-2)\,.
\end{equation}
In deriving (\ref{masval}) we use the commutation relations

\begin{eqnarray*}
&&{}[\,\heet{p}^2,p^A]=(1-d)p^A-2{\rm i}y^Ap^2+\gamma^A\heet{p}\,,
\\ [5pt]
&&{}[\,\heet{p}^2,y^A]=2{\rm i}p^A+(d-1)y^A+\gamma^A\heet{y}\,,
\\[5pt]
&&{} [\,\heet{p}^2,\theta^{AB}\gamma^B]
=y^A(-2{\rm i}\heet{p}-(d+1)\heet{y})+\gamma^A\,.
\end{eqnarray*}
Now comparing (\ref{masval}) and (\ref{masener}) it is seen
that there is a quadratic equation for $E_0$, whose solutions
read

\begin{equation}\label{enval}
E_0^{(1,2)}=
\left\{
\begin{array}{cccl}
d-s-1/2 &\hbox{ or }& s-1/2\quad &\hbox{ for } \Psi_{as}
\\ [7pt]
s+d-5/2& \hbox{ or }& -s+3/2& \hbox{ for }
\Psi_{sym}
\end{array}\right.
\end{equation}
As seen from (\ref{enval}) there exists an arbitrariness of
choosing $E_0$.  Thus we can conclude that the gauge invariance
by itself does not uniquely determine  the physical relevant
value of $E_0$ (in this regard the situation is similar to the
$d=4$ case, see \cite{F1}).  To choose a physical relevant
value of $E_0$ one can exploit the unitarity condition, that is,
(i) Hermiticity $(J^{AB})^\dagger=J^{AB}$; (ii)
the positive norm requirement. Making use of this unitarity
condition one can prove that the $E_0$ should satisfy the
inequalities $E_0\geq d-s-1/2$ for the case of $\Psi_{as}$
and $E_0\geq s+d-5/2$ for the case of $\Psi_{sym}$. Comparing
these inequalities with (\ref{enval}) we find the following
physical relevant values of $E_0$

\begin{equation}\label{envaltr}
E_0=
\left\{
\begin{array}{lll}
d-s-1/2\,,\qquad& 1<s\leq (d-2)/2\,,&\quad\hbox{ for } \Psi_{as}
\\ [7pt]
s+d-5/2\,,\qquad & s>0\,, &\quad\hbox{ for } \Psi_{sym}
\end{array}\right.
\end{equation}
where we recall domains of values of $s$.
Thus the analysis anti-de Sitter gauge transformations in
combination with unitarity enables us to fix the $E_0$, i.e.
gauge invariance and unitarity complement each other.
For the case of $d=4$ and $\Psi_{sym}$ the solution
$E_0=s+3/2$ is the well known result of \cite{FH}.
Note that our result for $E_0$ (\ref{envaltr}) cannot be
extended to cover the case of one-half spinor representation
($s=0$) because the $E_0$ obtained are relevant only for gauge
fields. This case should be considered in its own right and the
relevant value $E_0(\hbox{for } s=0)=(d-1)/2$ can be obtained by
using requirement of conformal invariance (see \cite{MET3}). The
case of $s=0$, $d=4$ has been investigated in \cite{DKS}.

Up to now we analysed second-order equations for $\Psi$.  Now we
would like to derive first-order equations. To do that we use
the relations

$$
\heet{p}=-{\rm i}\heet{y}\kappa,
\qquad
[\kappa,\heet{y}]=\heet{y}(-2\kappa-d)\,,
\qquad
\heet{y}^2=1\,,
$$

\begin{equation}\label{dirop}
\heet{p}^2=\kappa^2+d\kappa\,,
\qquad
\kappa\equiv \sigma^{AB}L^{AB}\,,
\end{equation}
and rewrite equation (\ref{eqmot2}) as follows:

$$
(\kappa+E_0+\frac{1}{2})(\kappa-E_0-\frac{1}{2}+d)|\Psi\rangle
=0
$$
Thus we could use the following two first-order equations of
motion:

\begin{equation}\label{eqmot3}
(\kappa-E_0-\frac{1}{2}+d)|\Psi\rangle
=0\,,
\end{equation}
\begin{equation}\label{eqmot4}
(\kappa+E_0+\frac{1}{2})|\tilde{\Psi}\rangle
=0\,.
\end{equation}
Due to relation

$$
(\kappa-E_0-\frac{1}{2}+d)|\Psi\rangle
=-\heet{y}(\kappa+E_0
+\frac{1}{2})\heet{y}|{\Psi}\rangle
$$
it is clear that $|\Psi\rangle$ and $|\tilde{\Psi}\rangle$ are
related by

$$
|\Psi\rangle=\heet{y}|\tilde{\Psi}\rangle
$$
i.e. equations (\ref{eqmot3}) and (\ref{eqmot4}) are
equivalent. We will use the equation of motion given by
(\ref{eqmot3}). Now we should verify  gauge invariance of
the first-order equation (\ref{eqmot3}) with respect to gauge
transformations (\ref{gaugetr2}). It turns out that the
invariance requirement of (\ref{eqmot3}) with respect to
(\ref{gaugetr2}) leads to

\begin{equation}\label{ssol}
S^{A_2\ldots A_s}=0\,.
\end{equation}
In deriving (\ref{ssol}) we use the commutation relations

\begin{eqnarray*}
{}& [\kappa,p^A]=\gamma^A\heet{p}-p^A\,,
\qquad
{} [\kappa,y^A]=\gamma^A\heet{y}-y^A\,,&
\\[6pt]
{}& [\kappa,\theta^{AB}\gamma^B]
=2{\rm i}p^A\heet{y}+(d-1)y^A\heet{y}+\gamma^A\,.&
\end{eqnarray*}
Thus the final form of gauge transformations is

\begin{equation}\label{gaugetr3}
\delta \Psi^{A_1\ldots A_s}
=\sigma(A)\left(
(p^{A_1}
\mp {\rm i}(s-1)y^{A_1})\Lambda^{A_2\ldots A_s}\right)
+\mbox{cyclic permutations}
\end{equation}
which can be rewritten in a simpler form

$$
\delta |\Psi\rangle
=a^A\left(p^A
\mp {\rm i}(s-1)y^A\right)|\Lambda\rangle
$$
or in terms of $E_0$ (see equation (\ref{envaltr})) as follows:

\begin{equation}\label{gaugetr5}
\delta |\Psi\rangle
=a^A\Bigl(p^A
+{\rm i}(E_0+\frac{3}{2}-d)y^A\Bigr)|\Lambda\rangle\,,
\end{equation}
where we use the notation

$$
|\Lambda\rangle=((s-1)!)^{-1}a^{A_2}\ldots a^{A_s}
\Lambda^{A_2\ldots A_s}|0\rangle\,.
$$
The  equations of motion for $\Lambda$ can be obtained
from (\ref{gammainv1}) and (\ref{ssol})

\begin{equation}\label{gpeqmot1}
(\kappa\pm (s-1))|\Lambda\rangle=0
\end{equation}
which can also be rewritten in terms of $E_0$

\begin{equation}\label{gpeqmot2}
(\kappa - E_0-\frac{3}{2}+d)|\Lambda\rangle=0\,.
\end{equation}
Making use of (\ref{ssol})  and (\ref{gpeqmot1}) one can
make sure that equation (\ref{mominv}) is also satisfied.

Thus we have constructed equations of motion (\ref{eqmot3}) which
respect gauge transformations (\ref{gaugetr5}), where the
gauge parameter fields $\Lambda$ satisfy the constraints
(\ref{gptrans})--(\ref{gptrac}) and equations of motion
(\ref{gpeqmot2}).  The relevant values of $E_0$ are given by
(\ref{envaltr}).  Note that the equations of motion are written
in terms of the operator $\kappa$ (see equation (\ref{dirop})).
The $\kappa$
was introduced in \cite{DIR} while constructing the equation of
motion for the field associated with the representation labeled
by $D(E_0,1/2)$. The $\kappa$ is expressible in terms of the
orbital part $L^{AB}$ (see equation (\ref{dirop})). Now we would
like to rewrite our equations of motion in terms of complete
angular momentum $J^{AB}$. In our opinion such a form is more
promising. To do that let us first multiply equations
(\ref{eqmot3}) and (\ref{gpeqmot2}) by $\heet{y}$. Then we use the
following equalities:

$$
y^Aa^BJ^{AB}=(ap)\mp {\rm i}(ya)(a\bar{a})+{\rm i}a^2(y\bar{a})
-\frac{{\rm i}}{2}a^A\theta^{AB}\gamma^B\heet{y}\,,
$$
$$
y^A\gamma^BJ^{AB}=-{\rm i}\heet{y}(\kappa+\frac{1}{2}d)
+y^A\gamma^BM_b^{AB}\,,
$$
where the expressions like $(ab)$ indicate the scalar product
$a^Ab^A$. Now making use of the equalities
$(a\bar{a})|\Psi\rangle=-s|\Psi\rangle$,
$(a\bar{a})|\Lambda\rangle=-(s-1)|\Lambda\rangle$ which tell us
that $\Psi$ and $\Lambda$ are tensors of rank
$s$ and $s-1$, respectively, and the equalities
$(y\bar{a})|\Psi\rangle=0$, $(y\bar{a})|\Lambda\rangle=0$ which
are the constraints (\ref{trans}) and (\ref{gptrans})
we rewrite the equations of motion (\ref{eqmot3}) as

\begin{equation}\label{eqmot5}
\Bigl(y^A\gamma^BJ^{AB}+{\rm i}(E_0-\frac{d-1}{2})\heet{y}
\Bigr)|\Psi\rangle=0\,,
\end{equation}
the gauge transformations (\ref{gaugetr5}) as

\begin{equation}\label{gaugetr6}
\delta|\Psi\rangle=y^A\Bigl(a^B
-\frac{a^C\theta^{CD}\gamma^D}{2E_0+3-d}\gamma^B\Bigr)
J^{AB}|\Lambda\rangle\,,
\end{equation}
and equations of motion for gauge parameter fields
(\ref{gpeqmot2}) as

\begin{equation}\label{gpeqmot3}
\Bigl(y^A\gamma^BJ^{AB}
+{\rm i}(E_0^\Lambda-\frac{d-1}{2})\heet{y}
\Bigr)|\Lambda\rangle=0\,,
\end{equation}
where we introduce lowest energy value for $\Lambda$:
$E_0^\Lambda=E_0+1$

\begin{equation}\label{gpenvaltr}
E_0^\Lambda=
\left\{
\begin{array}{ll}
d-s+1/2 \qquad&  \hbox{ for } \Lambda_{as}
\\ [7pt]
s+d-3/2 \qquad & \hbox{ for } \Lambda_{sym}
\end{array}\right.
\end{equation}
In deriving (\ref{gaugetr6}) we use the
equations of motion for $\Lambda$ (\ref{gpeqmot2}). Note also
that in deriving of (\ref{eqmot5})--(\ref{gpeqmot3}) it is
necessary to use the relevant values of $E_0$ given by
(\ref{envaltr}).

With the values for $E_0^\Lambda$ at hand we are ready to
provide an answer to the question: do the gauge parameter
fields meet the masslessness criteria? As expected, by
analogy with $d=4$ case the answer for the case of
$\Lambda_{sym}$ is negative. As for $\Lambda_{as}$ the answer is
positive. Indeed the inter-relations between of spin value $s$ and
energy value $E_0$ for massless fields are given by
(\ref{envaltr}).  Therefore the energy values for spin $s-1$
massless fields are obtainable by making substitutions
$s\rightarrow s-1$ in (\ref{envaltr}). Because of the relations

\begin{eqnarray}
\label{yesmas}
&&
E_0^\Lambda = E_0(s\rightarrow s-1)
\qquad \hbox{ for }\,\, \Lambda_{as}
\\ [5pt]
\label{nomas}
&&
E_0^\Lambda =\!\!\!\!\!/\,\,\, E_0(s\rightarrow s-1)
\qquad \hbox{ for }\,\, \Lambda_{sym}
\end{eqnarray}
we conclude that the $\Lambda_{as}$ is a massless field
while the $\Lambda_{sym}$ is a massive field. The result that
$\Lambda_{sym}$ is a massive field is a generalization of well
known results for $d=4$. To our knowledge the masslessness of
$\Lambda_{as}$ has not been referred to previously in the literature.
Note that this fact was not discovered in $d=4$ because for
this case there are no essentially totally  antisymmetric
representations, i.e. for the totally symmetric case one exhausts
all representations.

In conclusion, let us formulate the results of this letter. For
massless fermionic fields of arbitrary spins $s>0$
corresponding to totally (anti)sym\-metric
tensor-spinor representations of the $SO(d-1)$ compact subgroup
and in $d$-dimensional anti-de Sitter space
we have constructed: (i) free equations of motion (\ref{eqmot3}),
subsidiary conditions (\ref{trans})--(\ref{trac}),
corresponding gauge transformations (\ref{gaugetr5}), and
constraints (\ref{gptrans})--(\ref{gptrac}) and equations of motion
for the gauge parameter fields (\ref{gpeqmot2}); (ii)
the energy lowest values (\ref{envaltr}); (iii) the new
representation for equations of motion
(\ref{eqmot5}),(\ref{gpeqmot3}) and gauge transformations
(\ref{gaugetr6}) in terms of the generators of
the anti-de Sitter group $SO(d-1,2)$.
It is demonstrated that in contrast to the symmetric case
the gauge parameter of the antisymmetric massless field is
also a massless field
(see equations (\ref{yesmas}) and (\ref{nomas})).

This work was supported in part by INTAS, Grant No.94-2317,
by the Russian Foundation for Basic Research,
Grant No.96-01-01144, and by the NATO Linkage, Grant No.931717.

\end{document}